\documentclass[fleqn,twoside]{article}
\usepackage{espcrc2,epsfig}

\title{Non-Perturbative Renormalization and the Fermilab Action
\thanks{Talk presented in Lattice 2003.}}

\author{Huey-Wen Lin\address{Department of Physics, Columbia
University, New York, NY, 10027}}
\begin{document}


\begin{abstract}
We discuss the application of the regularization
independent (RI) scheme of Rome/Southampton to determine the
normalization of heavy quark operators non-perturbatively
using the Fermilab action.
\end{abstract}

\maketitle

\section{Introduction}

Heavy meson hadronic matrix elements play an essential role in
determining many of the parameters of the Standard Model, such as
the CKM matrix elements. Theoretically, lattice gauge theory
provides a means of computing these hadronic matrix elements with
better control of systematic
errors~\cite{previous_HQ_talks_01-02}. The heavy quark methods
that have been successfully used on the lattice are:
Sheikholeslami-Wohlert (SW) action, nonrelativistic QCD, heavy
quark effective theory (HQET) and Fermilab action.
Here we focus on the Fermilab action because it may
best control the systematic errors associated
with the large quark mass.

Renormalization of lattice operators is necessary in order to
obtain physical results from numerical simulations.
Non-perturbative renormalization (NPR) methods are attractive
since they avoid possibly large errors associated with lattice
perturbation theory.  This is especially important when domain
wall fermions are used for the light quarks.
Among various NPR methods, we decide to focus on RI/MOM scheme
NPR~\cite{RI_NPR}, because of the promising results~\cite{RBC_NPR}
reported by the RBC collaboration. We will use NPR to represent
the RI/MOM scheme NPR method in the rest of this paper.

Our goal is to adopt the RI/MOM scheme and to extend
its application to heavy quarks.  In particular we will show
how to generalize the usual NPR
``window restriction'' : $ \Lambda_{QCD}^2 << - p^2 << 1/a^2 $
to the kinematic region required by the Fermilab approach to
heavy quarks.


\section{Generalizing RI NPR to heavy quarks}

We need to find a new renormalization region, which meets three
requirements:
i) It permits evaluation in continuum perturbation theory.
Thus, it should involve non-exception external momenta, giving no
infrared singularities as light quark and gluon masses are set to zero.
ii) The Fermilab method remains valid, i.e. the region is not too far
from the heavy quark mass pole.
iii) It can be studied using a Euclidean, lattice Green's function.

\begin{figure}[!t]
\vskip -0.2in \epsfig{figure=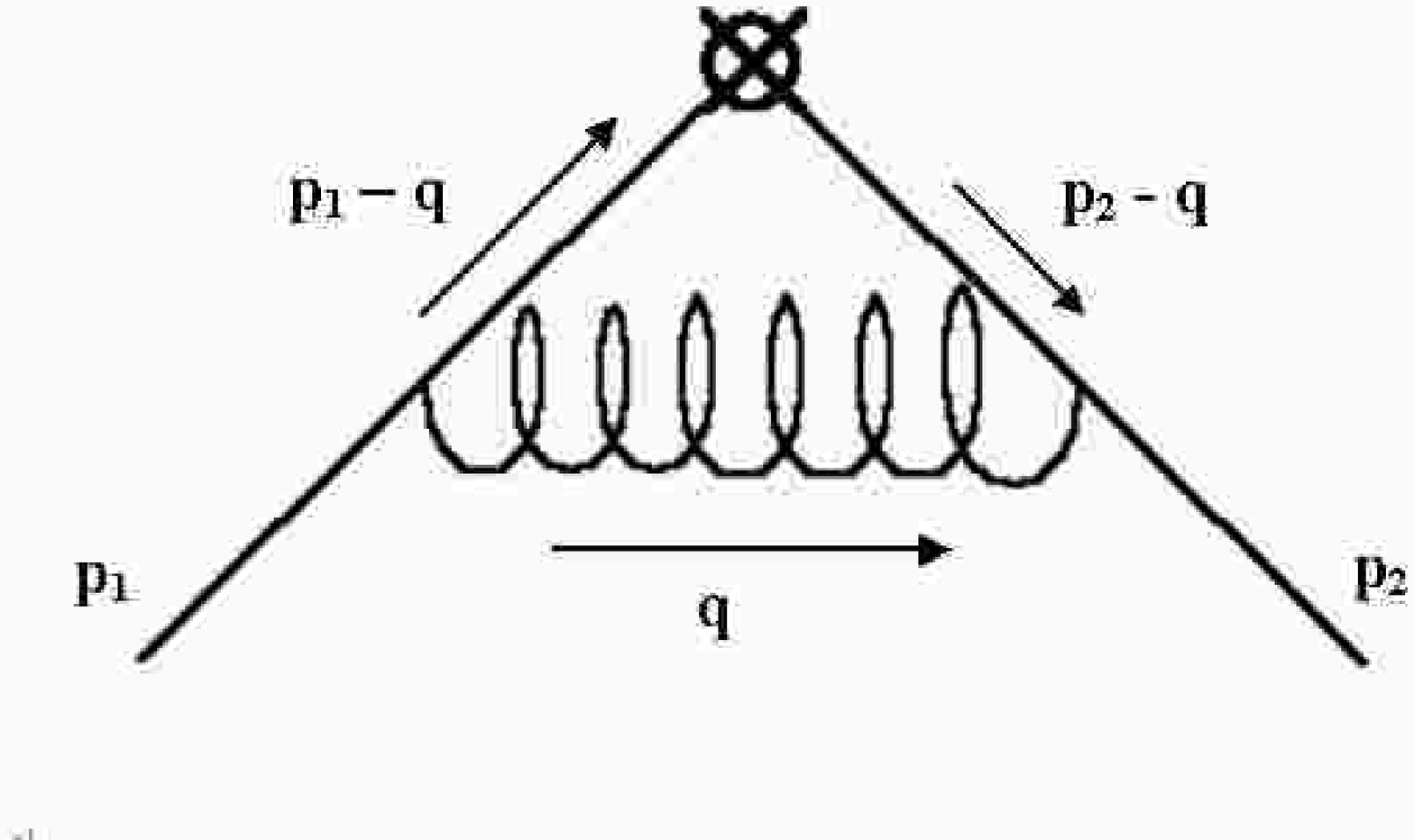,width=\columnwidth}
\vspace{-18mm} \caption{Example of a Feynman diagram in weak
interaction effective field theory. } \label{fig:feynman} \vskip
-7mm
\end{figure}

Let's take a look at a typical Feynman diagram with a one loop QCD
correction integral, for example, that shown in
Figure~\ref{fig:feynman}, with the integral
\begin{eqnarray}
\int \frac{d^4q}{ (2 \pi)^4} \frac {1}{q^2}
 \frac {i ({\ooalign{\hfil/\hfil\crcr$p_1$}}-{\ooalign{\hfil/\hfil\crcr$q$}}+m)}
 {(p_1-q)^2-m^2 } \frac {i ({\ooalign{\hfil/\hfil\crcr$p_2$}}
 -{\ooalign{\hfil/\hfil\crcr$q$}}+m)}{(p_2-q)^2-m^2 }.
\end{eqnarray}
Here q is the internal gluon momentum and $p_i$ is the net
external quark momentum flowing into $i^{th}$ internal quark line
within the loop.

The denominator of the $i^{th}$ quark line, with Minkowski
external momenta but Euclidean loop momentum is
$(\vec{p_i}-\vec{q})^2 -p_{i,0}^2+q_0^2 +m^2-2ip_{i,0}\cdot q_0$.
Since $q_0$ may be small, it doesn't guarantee that we can avoid
the pole on the $p_0$ axis during integration. Therefore,
additional
conditions have 
to be set to avoid singularities and permit the use of continuum
perturbation theory.  Specifically, we require that the absolute value
of each denominator be much larger than $\Lambda_{QCD}^2$.

Let's first check our formalism with a massless quark. We find
that $-p^2 = \mu^2 >> \Lambda_{QCD}^2$, which is consistent with
the the original NPR method for light quarks.

How about heavy quarks? We propose: $ m^2-p_0^2=
\mu^2>>\Lambda_{QCD}^2$. In order to keep Fermilab discretization
errors under control, we have to constrain the above quantity to
be much smaller than $1/a^2$ as well so that the heavy quark lines
remain nearly on-shell. Therefore, our ``window'' of
renormalization is $\Lambda_{QCD}^2<< m^2-p^2<< 1/a^2$.

\begin{figure}[!t]
\epsfig{figure=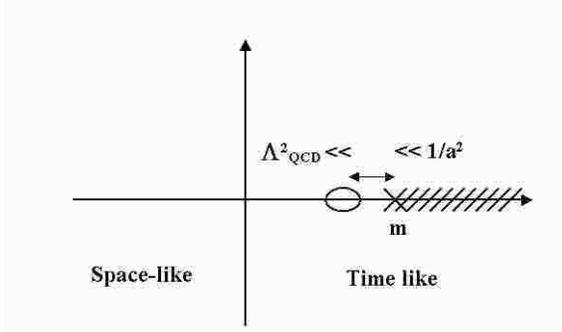,width=\columnwidth}
\vspace{-18mm} \caption{The complex-$p^2 = p_0^2-{\vec{p}}^2$ plane.}
\label{fig:q_sq_plane}
\vskip -7mm
\end{figure}

Figure~\ref{fig:q_sq_plane} is  a plot in complex $p^2$ plane. Our
choice of renormalization points are located within the small
region circled. The distance from this circled region to the mass
pole has to be much smaller than $\frac{1}{a^2}$ and much larger
$\Lambda_{QCD}^2$ for perturbation theory to be valid. This is
what I mean by ``slightly off-shell condition'' in the rest of the
paper.

Note that the region we choose is time-like, while the usual NPR's
choice is space-like. Why do we pick a time-like point, instead of
a space-like point? The constraint comes from the Fermilab action,
which is an on-shell O(a) improvement.
If we choose a deep, space-like point, the well-behaved
properties of Fermilab action would no longer be valid.

\section {Off-shell Fermilab action}

Recall ``heavy quark'' means the product of $m$ and
lattice spacing $a$ may be larger than or equal to 1. Under this
circumstance, it's natural to expect an improved action to break
the time-space symmetry. Here is the original Fermilab
action~\cite{Fermilab_action}:
\begin{eqnarray}
S &= & \sum_n \bar{\psi_n}\psi_n {} \nonumber\\
&-&\hskip
-0.15in\kappa_t\sum_n[\bar{\psi_n}(1-\gamma_0)\textsl{U}_{n,0}\psi_{n+\hat{0}}
 + \bar{\psi}_{n+\hat{0}}(1+\gamma_0)\textsl{U}_{n,0}^\dag\psi_n]   \nonumber\\
&-&\hskip -0.15in \kappa_s \sum_{n,i} [\bar{\psi}_n (r_s-\gamma_i)
\textsl{U}_{n,i}
+ \bar{\psi}_{n+\hat{i}} (r_s+\gamma_i) \textsl{U}_{n,i}^\dag \psi_n] \nonumber \\
&+&\hskip -0.15in
 \frac{i}{2} c_{B} \kappa_{s} \sum_{n;i,j,k} \varepsilon_{ijk}
\bar{\psi}_{n} \sigma_{ij}
B_{n;k} \psi_{n} \nonumber \\
&+&\hskip -0.15in ic_{E} \kappa_{s} \sum_{n;i} \bar{\psi}_{n}
\sigma_{0i} E_{n;i} \psi_{n}
\end{eqnarray}
Note that those coefficients are the functions of $ma$, and
are expected to remain finite for both small and large $ma$.

What happens when we go slightly off-shell? If the external quark
masses differ from the physical ones by a small amount $\delta
m$, with $a\delta m \ll 1$ then the resulting errors will be no
larger than the other discretization errors in the Fermilab
approach.

However, when we go off-shell, we have to include more terms to
make the O(a) improvement complete. (Why? because contact terms
and new non-gauge-invariant operators can appear.) First, we have
to add O(a) off-shell improvement terms in the action.
Fortunately, those terms which need to be added in the action can
be compensated by improving the quark
fields~\cite{Martinelli_offShell}. Therefore, the action itself
remains the same, including the well-behaved coefficients.

The improved quark fields will look similar to
those in ref~\cite{Martinelli_offShell}, but with broken
space-time symmetry.  Likewise for the composite operators.  The broken symmetry gives us at least
double the number coefficients to be determined non-perturbatively,
compared with the light quark cases. We are still exploring how
these coefficients may be determined in a more efficient way.

\section {Imposing RI NPR on the lattice}

To respect the ``slightly off-shell condition'',
we need to change the time-component Fourier transformation
to Laplace transformation.  Take the operator $O_{\Gamma}(x)=\bar{\psi}(x) \Gamma \psi (x)$ as an
example, for a general Dirac matrix $\Gamma$.
The non-amputated Green's function in momentum
space is:
\begin{equation}
G_O(pa) = \langle S_{\{U\}} (p \mid 0)\; \Gamma\;
\gamma_5 {S}_{\{U\}}^{\dagger} (p\mid 0)\gamma_5 \rangle|_{\{U\}}
\end{equation}
where $ S_i (p \mid y) = \int d^3 x \int dx_0 S_i (x \mid y)
e^{+ip\cdot x} e^{p_0x_0}$.

We calculate the amputated Green's function and apply the
projector onto $\hat{P}_O$ to define:
\begin{eqnarray}
\Gamma_O(pa) = \frac{1}{12}Tr[S(pa)^{-1} G_O(pa) S(pa)^{-1}
\hat{P}_O],
\end{eqnarray}
where the S(pa) is the gauge-averaged
propagator.  Following the usual NPR procedure, we apply the
renormalization condition:
\begin{eqnarray}
\langle p| O_{\Gamma}^{NPR}|p\rangle\mid_{m^2-p^2={\mu}^2} \;
  = \langle p| O_{\Gamma}|p\rangle_0,
\end{eqnarray}
where $\langle...\rangle_0$ represents tree-level value.
Thus, imposing
$Z_OZ_{\psi}^{-1}\Gamma_O|_{m^2-p^2={\mu}^2} \; =1 $ will determine the renormalization
factor $Z_O(\mu a, g(a))$.

\section {Matching to the $\overline{MS}$ scheme }

Our lattice NPR calculations are done in the RI renormalization
scheme. We need continuum RI renormalization calculations to
match to physical values given in the continuum $\overline{MS}$ scheme.
The previous matching calculations done at $-p^2 = \mu^2$, while
taking $m \rightarrow 0$ must be extended to our new scheme.

Let's take a look at the simplest example: the calculation of
$Z_q$ and $Z_m$.  Our new renormalization conditions become:
\begin{eqnarray}
\frac{1}{48}Z_q Tr[\gamma_{\mu}
\frac{\partial({\ooalign{\hfil/\hfil\crcr$p$}}\bar{\Sigma}_1(p))}{\partial
p_{\mu}}]_{m^2-p^2 = \mu^2} = 1
\end{eqnarray}
\begin{eqnarray}
\frac{1}{12} Z_q Z_m Tr[\bar{\Sigma}_2(p)]_{m^2-p^2= \mu^2} = 1
\end{eqnarray}
where $iS^{-1}(p) =  Z_q
[{\ooalign{\hfil/\hfil\crcr$p$}}\bar{\Sigma}_1(p) - Z_m m
\bar{\Sigma}_2 (p)] $.

The calculations are done in the dimensional regularization
scheme, where D is 4-2$\epsilon$. Here are the matching factors
for $Z_m$ and $Z_q$ in the RI and $\overline{MS}$ schemes:
\begin{eqnarray}
\frac{Z_m^{RI} }{ Z_m^{\overline{MS}}}  \hskip -0.1in &=&
1 - \frac{\alpha_s}{4\pi}\frac{N_c^2 -1}{2 N_c}\{ \frac{8 + 3\xi}{2} -
3\xi \frac{m^2}{m^2 - \mu^2}\nonumber \\
&-& \hskip -0.15in [ ( 3 + \xi ) \frac{m^2}{m^2 - \mu^2}- 3 \xi
\frac{m^4}{(m^2 - \mu^2)^2}]ln \frac{m^2}{\mu^2}\nonumber \\
&+& \hskip -0.15in ln\frac{\mu'\;^2}{\mu^2}( 3 + 2\xi) \}  + O(\alpha_s^2)
\end{eqnarray}
\begin{eqnarray}
\frac{Z_q^{RI} }{ Z_q^{\overline{MS}}} \hskip -0.1in
&=& 1 - \frac{\alpha_s}{4\pi} \frac{N_c^2 -1}{2 N_c} \xi (
\frac{1}{2} + 3 \frac{m^2}{m^2 - \mu^2}{}\nonumber\\
-&3&\hskip -0.15in \frac{m^4}{(m^2 - \mu^2)^2} ln
\frac{m^2}{\mu^2}-ln\frac{\mu'\;^2}{\mu^2}) + O(\alpha_s^2),
\end{eqnarray}
where $\mu'$ is defined as $g_0 = Z_g g \mu'\;^{\epsilon}$.

\section{Conclusion}

In this paper, we present a first NPR renormalization proposal
for relativistic heavy quarks. We propose a new renormalization
point: $\Lambda_{QCD}^2<< m^2-p^2<< 1/a^2$, in a slightly
off-shell region.   We must add off-shell improvement
terms and determine the corresponding coefficients as well as
evaluating Laplace transformations in the time.

The author would like to thank each member of the RBC
collaboration for their assistance. Special thanks go to N.
H. Christ for inspiration and constant discussions on various
topics.


\begin{thebibliography}{1}
\expandafter\ifx\csname bibnamefont\endcsname\relax
  \def\bibnamefont#1{#1}\fi
\expandafter\ifx\csname bibfnamefont\endcsname\relax
  \def\bibfnamefont#1{#1}\fi
\expandafter\ifx\csname url\endcsname\relax
  \def\url#1{\texttt{#1}}\fi
\expandafter\ifx\csname urlprefix\endcsname\relax\def\urlprefix{URL }\fi
\expandafter\ifx\csname bibinfo\endcsname\relax \def\bibinfo#1#2{#2}\fi
\expandafter\ifx\csname eprint\endcsname\relax \def\eprint#1{#1}\fi



\bibitem{previous_HQ_talks_01-02}
\bibinfo{author}{\bibfnamefont{S. M.}~\bibnamefont{Ryan}},
  \bibinfo{journal}{Nucl.Phys.Proc.Suppl.} \textbf{\bibinfo{volume}{106}},
  \bibinfo{pages}{86} (\bibinfo{year}{2002}),
  \eprint{hep-lat/0111010};
\bibinfo{author}{\bibfnamefont{N.}~\bibnamefont{Yamada }},
  \eprint{hep-lat/0210035}.

\bibitem{RI_NPR}
\bibinfo{author}{\bibfnamefont{G.}~\bibnamefont{Martinelli }}\emph{et~al.},
  \bibinfo{journal}{Nucl.Phys.} \textbf{\bibinfo{volume}{B445}},
  \bibinfo{pages}{81} (\bibinfo{year}{1995}),
  \eprint{hep-lat/9411010};
\bibinfo{author}{\bibfnamefont{V.}~\bibnamefont{Gimenez }}\emph{et~al.},
  \bibinfo{journal}{Nucl.Phys.} \textbf{\bibinfo{volume}{B531}},
  \bibinfo{pages}{429} (\bibinfo{year}{1998}),
  \eprint{hep-lat/9806006};
\bibinfo{author}{\bibfnamefont{A.}~\bibnamefont{Doninir }}\emph{et~al.},
  \bibinfo{journal}{Eur.Phys.J.} \textbf{\bibinfo{volume}{C10}},
  \bibinfo{pages}{121} (\bibinfo{year}{1999}),
  \eprint{hep-lat/9902030}.

\bibitem{RBC_NPR}
\bibinfo{author}{\bibfnamefont{T.}~\bibnamefont{Blum }} \emph{et~al.}
  (\bibinfo{collaboration}{RBC}),  (\bibinfo{year}{2001}),
  \eprint{hep-lat/0110075};
  \bibinfo{author}{\bibfnamefont{T.}~\bibnamefont{Blum }} \emph{et~al.}
  (\bibinfo{collaboration}{RBC}), \bibinfo{journal}{Phys.Rev.}
  \textbf{\bibinfo{volume}{D66}},
  \bibinfo{pages}{014504} (\bibinfo{year}{2002}),
  \eprint{hep-lat/0102005}.


\bibitem{Fermilab_action}
\bibinfo{author}{\bibfnamefont{A. X.}~\bibnamefont{El-Khadra }} \emph{et~al.}
    , \bibinfo{journal}{Phys.Rev.}
  \textbf{\bibinfo{volume}{D55}}, \bibinfo{pages}{3933} (\bibinfo{year}{1999}),
  \eprint{hep-lat/9604004}.

\bibitem{Martinelli_offShell}
\bibinfo{author}{\bibfnamefont{G.}~\bibnamefont{Martinelli }}\emph{et~al.},
  \bibinfo{journal}{Nucl.Phys.} \textbf{\bibinfo{volume}{B611}},
  \bibinfo{pages}{311} (\bibinfo{year}{2001}),
  \eprint{hep-lat/0106003}.



\end{thebibliography}

\end{document}